\documentclass{article}

\usepackage{amsmath}
\usepackage{natbib}
\usepackage{soul}
\usepackage{color}
\usepackage{setspace}
\usepackage{amsthm}
\usepackage{amssymb}
\usepackage{geometry}

\newcommand{\var}{\mathbf{var\,}}
\newcommand{\E}{\mathbf{E}}
\renewcommand{\P}{\mathbf{P}}
\newcommand{\I}{\mathbb{I}}
\newcommand{\Z}{\mathbb{Z}}
\newcommand{\R}{\mathbb{R}}

\newcommand{\revision}[1]{{\color{red}{#1}}}
\renewcommand{\revision}[1]{#1}

\newlength{\zero}
\newtheorem{thm}{Theorem}

\numberwithin{equation}{section}

\title{Exact Estimation for Markov Chain Equilibrium Expectations} % insert title - use \\ if it requires more than one line.
\author{
    Peter W. Glynn \\
    Stanford University\\
	Stanford, CA, 94305, USA\\\\
    Chang-han Rhee\\
	Georgia Institute of Technology\\
	Atlanta, GA, 30332, USA\\\\
}
\date{\today}

\begin{document}

\maketitle
% \authorone[Affiliation]{Peter W. Glynn and Chang-han Rhee} % Affiliation is just the name of your university or institution
% \authorone[Stanford University]{Peter W. Glynn} 
% \authortwo[Georgia Institute of Technology]{Chang-han Rhee} % Affiliation is just the name of your university or institution

% \addressone{Full address} % Your postal address goes here.
\onehalfspacing
\begin{abstract}
\noindent
We introduce a new class of Monte Carlo methods, which we call exact estimation algorithms. Such algorithms provide unbiased estimators for equilibrium expectations associated with real-valued functionals defined on a Markov chain. We provide easily implemented algorithms for the class of positive Harris recurrent Markov chains, and for chains that are contracting on average. We further argue that exact estimation in the Markov chain setting provides a significant theoretical relaxation relative to exact simulation methods.
\end{abstract}

% \keywords{} % insert keywords separated by a semicolon

% \ams{}{} % insert the primary Maths Subject Classification number in the first bracket
         % and the secondary ams number(s) in the second bracket
         % e.g. \ams{60E20}{49G03;49F10}

%%%%%%%%%%%%%%%%%%%%%%%%%%%%%%%%%%%%%%%%%%%%%%%%%%%%%%%%%%%%%%%%%%%%%%%%%%%%
%
%
% Section 1. Introduction
%
%
%%%%%%%%%%%%%%%%%%%%%%%%%%%%%%%%%%%%%%%%%%%%%%%%%%%%%%%%%%%%%%%%%%%%%%%%%%%%
\section{Introduction}\label{sec:1}

A key advance in the development of Monte Carlo algorithms for Markov chains has been the introduction of what are known as \emph{exact simulation} or \emph{perfect simulation} algorithms for equilibrium distributions of such stochastic processes. 
By assuming suitable structure on the underlying Markov chain, one can construct algorithms that draw samples perfectly from the equilibrium distribution of the Markov chain, based only on an ability to generate sample paths of the Markov chain from an arbitrary state. 
In particular, such algorithms have been developed for finite state Markov chains \citep{Propp1996}, uniformly recurrent Markov chains \citep{Asmussen1992}, certain stochastically monotone Markov chains \citep{Propp1996,Corcoran2001}, some queueing models \citep{Ensor2000,Blanchet2014}, and various sub-classes of Harris recurrent Markov chains \citep{Kendall2004,Connor2007}. 
While the idea is powerful, it apparently requires exploiting significant structure within the Markov chain itself. 
In particular, no universal and practically implementable such perfect sampler has been constructed for Harris recurrent Markov chains, or even for countable state positive recurrent Markov chains. 
In fact, we shall argue below in Section~\ref{sec:2} that exact simulation is inherently restrictive, in the sense that such algorithms can typically be constructed only for Markov chains that are $\phi$-irreducible. 

In this paper, we relax the algorithmic formulation so as to require only that the algorithm output unbiased estimators for equilibrium expectations, rather than to demand (as in exact simulation) that such an unbiased estimator necessarily is constructed from an exact sample from the equilibrium distribution.
To differentiate this new class of algorithms from exact simulation algorithms, we shall refer to them as \emph{exact estimation} algorithms. 
Our exact estimation algorithms exploit a recent idea of \cite{Rhee2013} that shows how unbiased estimators can often be constructed from a sequence of biased approximations; see also \cite{McLeish2011} and \cite{Rhee2012}. 
As we shall see below, this new class of algorithms can provide unbiased estimators for equilibrium expectations for any positive recurrent Harris chain. 
In fact, we shall see that exact estimation algorithms can even be developed for non $\phi$-irreducible Markov chains, provided that the Markov chain is ``contractive'' in a certain sense and the equilibrium expectation involves a suitably Lipschitz functional (Theorem 1). 
This makes clear that exact estimation is indeed a significant relaxation of exact simulation.

One key additional feature of our exact estimation algorithms is that unlike most existing exact simulation methods, our algorithms do not involve explicitly simulating paths from multiple initial conditions, nor do they require monotonicity. 
Furthermore, our exact estimation procedures can easily be implemented with a minimal need to store sample paths. 
As a consequence, our proposed exact estimation methods are relatively straightforward to implement.

Our paper is organized as follows. 
In Section~\ref{sec:2}, we illustrate exact estimation in the setting of contractive Markov chains, and exploit the fact that Markov chains can, in great generality, be viewed as a sequence of random iterated functions. 
Section~\ref{sec:3} develops exact estimation algorithms in the context of Harris recurrent Markov chains, thereby establishing that our proposed relaxation is indeed a generalization of exact simulation.
Section~\ref{sec:4} proves a variant of the Glivenko-Cantelli theorem for our newly developed estimator, and Section~\ref{sec:5} provides a brief computational discussion.
%%%%%%%%%%%%%%%%%%%%%%%%%%%%%%%%%%%%%%%%%%%%%%%%%%%%%%%%%%%%%%%%%%%%%%%%%%%%
%
%
% Section 2. Exact Estimation for Contracting Markov Chains
%
%
%%%%%%%%%%%%%%%%%%%%%%%%%%%%%%%%%%%%%%%%%%%%%%%%%%%%%%%%%%%%%%%%%%%%%%%%%%%%
\section{Exact Estimation for Contracting Markov Chains}\label{sec:2}
Given an $S$-valued Markov chain $X=(X_n: n \geq 0)$, we wish to develop exact estimation algorithms for computing $\E f(X_\infty)$, where $f$ is real-valued and $X_\infty$ has the equilibrium distribution $\pi$ of $X$ (assumed to exist uniquely). 
We start by briefly describing and (slightly) generalizing the algorithms and analysis presented in \cite{Rhee2012,Rhee2013} and \cite{McLeish2011}, following related earlier work by \cite{Rychlik1990,Rychlik1995}.

Suppose that we wish to compute $\E Y$. 
We assume that we have available to us a sequence $(Y_k: k\geq 0)$ of approximations for which 
	\begin{equation}\label{eq:2.1}
	\E Y_k \to \E Y
	\end{equation}
as $k\to \infty$. If $(\Delta_k: k\geq 0)$ is a sequence of random variables (rv's) for which $\E \Delta_k = \E(Y_k - Y_{k-1})$ for $k\geq 0$ (with $Y_{-1} \triangleq 0$) and 
	\begin{equation}\label{eq:2.2}
	\E \sum_{k=0}^\infty |\Delta_k| < \infty,
	\end{equation}
then it is easy to verify that
	\begin{equation}\label{eq:2.3}
	Z = \sum_{k=0}^N \frac{\Delta_k}{\P(N\geq k)}
	\end{equation}
is an unbiased estimator for $\E Y$, provided that $N$ is a $\Z_+$-valued rv independent of $(\Delta_k: k \geq 0)$. 
Of course, one implication of (\ref{eq:2.2}) is that $\Delta_k \Rightarrow 0$ as $k\to \infty$, where $\Rightarrow$ denotes weak convergence. 
Note that in our Markov chain setting, the most natural choice of approximating sequence $(Y_k: k\geq 0)$ is to choose $Y_k = f(X_k)$ (with $Y = f(X_\infty)$). 
However, the obvious choice for $\Delta_k$, namely $\Delta_k = f(X_k ) - f(X_{k-1})$, then fails to satisfy $\Delta_k \Rightarrow 0$ as $k\to \infty$. 
Thus, the key to the development of an exact estimation algorithm for $X$ is the construction of a computationally implementable ``coupling'' between $X_{k-1}$ and $X_k$ that forces $\Delta_k$ to converge to zero (hopefully rapidly). 

We now illustrate one potential coupling that can be applied in the setting of ``contractive'' Markov chains. 
We presume (for the purpose of this section) that $S$ is a complete separable metric space with metric $\rho:S\times S\to \R_+$ and that $X$ can be represented in terms of a sequence of independent and identically distributed (iid) random functions $(\varphi_i: i\geq 1)$, independent of $X_0$, so that
	\begin{equation*}
	X_i = \varphi_i(X_{i-1})
	\end{equation*}
for $i \geq 1$. 
In particular, conditional on $X_0 = x \in S$, $X_n = (\varphi_n \circ \varphi_{n-1}\circ \cdots \circ \varphi_1)(x)$. 
An obvious means of coupling $X_{n-1}$ and $X_n$ is then to set 
	\begin{equation*}
	\tilde X_{n-1} = (\varphi_n\circ \varphi_{n-1}\circ \cdots \circ \varphi_2)(x)
	\end{equation*}
for $n\geq 2$, with $\tilde X_0 \triangleq x$. 
Clearly, $\tilde X_{n-1} \stackrel{\mathcal D}{=} X_{n-1}$ for $n\geq 1$, where $\stackrel{\mathcal D}{=}$ denotes ``equality in distribution.'' 

For $y, z \in S$, set 
	\begin{equation*}
	r(y,z) = \E \rho^2(\varphi_1(y), \varphi_1(z))
	\end{equation*}
and assume that there exists $b<1$ for which 
	\begin{equation}\label{eq:2.4}
	r(y,z) \leq b\rho^2(y,z)
	\end{equation}
for $y, z \in S$, so that $X$ is ``contractive on average.''
Suppose further that $f$ is Lipschitz with respect to the metric $\rho$, so that there exists $\kappa<\infty$ for which 
	\begin{equation}\label{eq:2.5}
	|f(y) - f(z)| \leq \kappa\rho(y,z)
	\end{equation}
for $y,z \in S$. 
% \revision{\st{If we now set $\Delta_k = f(X_k) - f(\tilde X_{k-1})$, e}} 
Evidently,
	\begin{align*}
	\E\Delta_k^2 
	&\leq \kappa^2 \E \rho^2(X_k, \tilde X_{k-1})\\
	&= \kappa^2 \E \rho^2(\varphi_k(X_{k-1}), \varphi_k(\tilde X_{k-2}))\\
	&\leq \kappa^2 b \E \rho^2(X_{k-1}, \tilde X_{k-2})\\
	&\leq \ldots \leq \kappa^2 b^{k-1} \E \rho^2(X_1, x).
	\end{align*}
Hence, if 
	\begin{equation}\label{eq:2.6}
	\E \rho^2(\varphi_1(x), x) < \infty
	\end{equation}
for $x\in S$, it follows that $\E \Delta_k^2 \to 0$ geometrically fast, so that (\ref{eq:2.2}) holds. 
Also, Theorem~\ref{thm:1} of \cite{Diaconis1999} applies, in the presence of (\ref{eq:2.4}) and (\ref{eq:2.6}), so that $X$ has a unique stationary distribution $\pi$. 
In fact, their proof makes clear that $\E f(X_n)\to\E f(X_\infty)$ geometrically fast (where $X_\infty$ has distribution $\pi$) when $f$ is Lipschitz. 
Consequently, (\ref{eq:2.1}) is valid, thereby proving that $Z$ is an unbiased estimator for $\E f(X_\infty)$. 

But more can be said. 
Note that 
	\begin{equation}\label{eq:*}
	\E Z^2  = \sum_{k=0}^\infty \frac{\E \Delta_k^2 + 2\sum_{j=k+1}^\infty \E \Delta_k \Delta_j}{\P(N\geq k)}.
	\end{equation}
By virtue of the Cauchy-Schwarz inequality, it follows that $\var Z < \infty$, provided that we choose the distribution for $N$ so that
	\begin{equation}\label{eq:2.7}
	\sum_{k=0}^\infty b^k / P(N\geq k) < \infty
	\end{equation}
Under condition (\ref{eq:2.7}), the central limit theorem (CLT) asserts that if we generate iid copies $Z_1, Z_2, \ldots$ of $Z$ and form the sample mean $\bar Z_n = (Z_1 + \cdots + Z_n)/n$, then $\bar Z_n$ converges weakly to $\E f(X_\infty)$ at rate $n^{-1/2}$ in the number of samples $n$ that are generated. 

Of course, the amount of computer time needed to generate each $Z_i$ could be excessive. 
To take this effect into account, we let $\xi_i$ be the computer time needed to generate $Z_i$. 
In view of the fact that computing $Z$ requires generating $\varphi_1, \varphi_2, \ldots, \varphi_N$, it seems natural to assess the computer time as being equal to $N$. 
Hence, we put $\xi_i = N_i$, where $N_i$ is the corresponding ``randomization'' rv $N$ associated with $Z_i$. 
If $\Gamma(c)$ is the number of $Z_i$'s generated in $c$ units of computer time, $\Gamma(c) = \max\{k\geq0: \xi_1 + \cdots + \xi_k \leq c\}$ and the estimator for $\E f(X_\infty)$ available after expending $c$ units of computer time is $\bar Z_{\Gamma(c)}$. 
If 
	\begin{equation}\label{eq:2.9}
	\E \xi_i = \E N = \sum_{k=0}^\infty \P(N\geq k) < \infty
	\end{equation}
and $\var Z<\infty$, it is well known (see, for example, \cite{Glynn1992}) that 
	\begin{equation*}
	c^{1/2} (\bar Z_{\Gamma(c)} - \E f(X_\infty)) \Rightarrow \sqrt{\E N \cdot \var Z} \ N(0,1)
	\end{equation*}
as $c\to\infty$, where $N(0,1)$ denotes a normal rv with mean $0$ and variance $1$. 

We summarize our discussion with the following result, which establishes that exact estimation algorithm exhibiting ``square root'' convergence rates can be obtained for any suitably contractive chain and Lipschitz function $f$. 

\begin{thm} \label{thm:1}
Assume (\ref{eq:2.4}) - (\ref{eq:2.6}). 
If the distribution of $N$ is chosen so that (\ref{eq:2.7}) and (\ref{eq:2.8}) hold (e.g. $\P(N\geq k) = ck^{-\alpha}$ for $\alpha > 1$), then 
	\begin{equation*}
	c^{1/2} (\bar Z_{\Gamma(c)} - \E f(X_\infty))\Rightarrow\sqrt{\E N \cdot \var Z} \ N(0,1)
	\end{equation*}
as $c \to \infty$.
\end{thm}

In view of Theorem~\ref{thm:1}, a natural question arises as to the optimal choice for the distribution of $N$. 
According to Proposition~1 of \cite{Rhee2013} the optimal choice is to set 
	\begin{equation}\label{eq:2.8}
	\P(N\geq k) = \left(\frac{\E\Delta_k^2 + 2\sum_{j=k+1}^\infty \E \Delta_k \Delta_j}{\E \Delta_0^2 + 2\sum_{j=1}^\infty \E \Delta_0 \Delta_j}\right)^{1/2}
	\end{equation}
for $k\geq 0$, provided that the right-hand side is non-\revision{increasing}; see Theorem~3 of \cite{Rhee2013} for details of the form of the optimal distribution when the right-hand side fails to be non-\revision{increasing}. Given this result and the geometric decay of the $\E \Delta_k^2$, it therefore seems reasonable to expect that requiring the tail of $N$ to be geometric will often be a good choice in this setting.

We turn next to a slightly different implementation of our coupling idea in the $S$-valued metric space contractive setting. Given the independence of $N$ from the $\varphi_i$'s, an alternative coupling for $(X_{i-1}, X_i)$ is to set 
	\begin{equation*}
	X_j^* = (\varphi_N \circ \varphi_{N-1}\circ \cdots \circ \varphi_{N-j+1}) (x)
	\end{equation*}
for $0 \leq j \leq N$. Clearly, conditional on $N$, $X_j^* \stackrel{\mathcal D}{=} X_j$ and $\Delta_j^* \triangleq X_j^* - X_{j-1}^* \stackrel{\mathcal D}{=} \Delta_j$ for $j\geq 0$. 
Because $\E (\Delta_j^*)^2 = \E \Delta_j^2$ for $j\geq 0$, the same argument as for $Z$ shows that 
	\begin{equation*}
	Z^* = \sum_{j=0}^N \frac{\Delta_j^*}{\P(N\geq j)}
	\end{equation*}
is unbiased for $\E f(X_\infty)$. 
Furthermore, the estimator for $\E f(X_\infty)$ corresponding to computing a sample average of iid copies of $Z^*$ satisfies Theorem~\ref{thm:1}, under the conditions (\ref{eq:2.4})-(\ref{eq:2.8}). 

The estimator based on $Z^*$ is slightly more complicated to implement, because $X_i^*$ can not be recursively computed from $X_{i-1}^*$ in this setting (whereas $(X_i, \tilde X_i)$ can be recursively computed from $(X_{i-1}, \tilde X_{i-1})$). 
Of course, the estimator based on $Z^*$ will have a different variance than does $Z$, because $\E \Delta_k \Delta_j \neq \E \Delta_k^*\Delta_j^*$ for $k<j$. 
In particular, while all four of the quantities $f(X_k^*)$, $f(X_{k-1}^*)$, $f(X_j^*)$, $f(X_{j-1}^*)$ appearing in $\E \Delta_k^*\Delta_j^*$ will be close to one another when $k$ is large, $f(X_j)- f(X_k)$ will exhibit significant variability, regardless of the magnitude of $k$. 
%%
% To gain some insight into this issue, we again suppose that the $\varphi_i$'s and $f$ are smooth, with $S\subseteq \R^d$. Note that $((X_j, \tilde X_j): j \geq 0)$ is a contractive Markov chain, so $\Delta_k \Rightarrow \Delta_\infty$ as $k\to \infty$. Furthermore, also under (\ref{eq:2.4})-(\ref{eq:2.6}), the $\Delta_j$'s are mixing, so that $\E \Delta_k \Delta_j\to \E \Delta_k \Delta_\infty$ as $j\to \infty$. On the other hand, for $m_1\ll m_2 \ll k \leq N$, 
% 	\begin{equation*}
% 	\Delta_k^*\approx \nabla f(X_N^*)R^*_N R_{N-1}^* \cdots R_{N-m_1}^*(X_{m_1:m_2}^* - X_{m_1:m_2-1})
% 	\end{equation*}
% where
% 	\begin{align*}
% 	R^*_{N-i} 
% 	&= R_{N-i}((\varphi_{N-i-1}\circ \cdots \circ \varphi_{N-m_2+1})(x)),\\
% 	X_{m_1:m_2}^*
% 	&= (\varphi_{N-m_1-1}\circ \cdots \circ \varphi_{n-m_2+1})(x).
% 	\end{align*}
% Thus, for $k$ large and $k\leq j$, $\E \Delta_k^*\Delta_j^*\approx \E (\Delta_k^*)^2 = \E (\Delta_\infty^*)^2 = \E \Delta_\infty^2$, so $\E \Delta_k^*\Delta_j^*$ is larger asymptotically then $\E\Delta_k\Delta_j \approx (\E \Delta_\infty)^2$. Hence, this crude analysis suggests that basing estimation on $Z$ will often be preferable to use of $Z^*$. 

We close this section by noting that construction of exact simulation algorithm typically requires that the underlying Markov chain be $\phi$-irreducible. 
Recall that an $S$-valued Markov chain $X=(X_n: n\geq 0)$ is \emph{$\phi$-irreducible} if there exists a $\sigma$-finite (non-negative) measure $\phi$ such that whenever $\phi(A)>0$ for some (measurable) $A$,
$$R(x,A) \triangleq \sum_{n=0}^\infty 2^{-n} \P_x(X_n\in A) > 0$$
for all $x\in S$, where $\P_x(\cdot) \triangleq \P(\cdot|X_0 = x)$. 
Equivalently,
$$\phi(\cdot) \ll R(x,\cdot)$$
for each $x\in S$, where $\ll$ denotes ``is absolutely continuous with respect to.''

A typical exact simulation algorithm involves simulating $X$ from multiple initial states $x_1, x_2, \cdots$, thereby yielding a family of random variables (rv's) $(X_{ij}: i \geq 1, j\geq 0)$ such that 
$$ \P((X_{ij}: j \geq 0) \in \cdot) = 
\P((X_j: j\geq 0)\in \cdot | X_0 = x_i)$$
for $i\geq 1$; the exact simulation algorithm then outputs $X_{IJ}$ for some appropriately chosen pair of rv's $(I,J).$ Exact simulation demands that if $X$ has an equilibrium distribution $\pi(\cdot)$, then 
$$ \P(X_{IJ} \in \cdot) = \pi(\cdot).$$
The probability $\P(X_{IJ}\in \cdot)$ is mutually absolutely continuous with respect to $\E 2^{-I-J} \I(X_{IJ} \in \cdot).$ But 
\begin{align*}
\E 2^{-I-J} \I(X_{IJ} \in \cdot)  
&= \sum_{i=1}^\infty \sum_{j=0}^\infty 2^{-i-j} \P(X_{ij}\in \cdot, I=i, J=j)\\
&\leq \sum_{i=1}^\infty \sum_{j=0}^\infty 2^{-i-j} \P(X_{ij}\in \cdot)\\
&= \sum_{i=1}^\infty 2^{-i}R(x_i, \cdot).
\end{align*}
Hence, it follows that
\begin{equation}\label{eq:2.99}
\pi(\cdot) \ll \sum_{i=1}^\infty 2^{-i} R(x_i, \cdot).
\end{equation}
Thus, the existence of an exact simulation algorithm requires that one have a priori knowledge of a set of states $x_1, x_2, \cdots$ satisfying (\ref{eq:2.99}). 
Without additional structure on the chain, the only way to guarantee this is to require that 
$$\pi(\cdot) \ll R(x,\cdot)$$
for each $x\in S$. 
In other words, $X$ must be $\phi$-irreducible, with $\phi = \pi$. 
(Note that when a $\phi$-irreducible Markov chain has a stationary distribution, one choice for $\phi$ is always $\phi = \pi$.) Consequently, $\phi$-irreducibility and exact simulation are tightly connected concepts.

%%%%%%%%%%%%%%%%%%%%%%%%%%%%%%%%%%%%%%%%%%%%%%%%%%%%%%%%%%%%%%%%%%%%%%%%%%%%
%
%
% Section 3. Exact Estimation for Harris Recurrent Markov Chains
%
%
%%%%%%%%%%%%%%%%%%%%%%%%%%%%%%%%%%%%%%%%%%%%%%%%%%%%%%%%%%%%%%%%%%%%%%%%%%%%
\section{Exact Estimation for Harris Recurrent Markov Chains}\label{sec:3}
In this section, we will establish that exact estimation algorithms can be constructed for any positive recurrent Harris chain, under the assumption that $S$ is a separable metric space. 
In the presence of such separability, it is well known that there exists a so-called small set $A$ i.e., there exists $m\geq 1$, a probability $\nu$, and $\lambda > 0$, for which 
	\begin{equation}\label{eq:3.1}
	\P(X_m \in \cdot | X_0 = x) \geq \lambda \nu(\cdot)
	\end{equation}
for all $x \in A$. 
Given (\ref{eq:3.1}), we can express the $m$-step transition probability on $A$ via the mixture representation
	\begin{equation}\label{eq:3.2}
	\P(X_m\in \cdot |X_0 = x) = \lambda \nu(\cdot) + (1-\lambda)Q(x,\cdot),
	\end{equation}
where $Q(x,\cdot)$ is a probability on $S$ for each $x\in A$. 
In view of (\ref{eq:3.2}), we can construct regeneration times $T(1),T(2),\ldots$ for $X$ by first running the chain until it hits $A$. 
Once it hits $A$ at time $T$ (say), we distribute the chain at time $T+m$ according to $\nu$ with probability $\lambda$ and according to $Q(X_T, \cdot)$ with probability $1-\lambda$; we then ``condition in'' the values of $X_{T+1}$, \ldots, $X_{T+m-1}$, conditional on $X_T$ and $X_{T+m}$. 
If we succeed in distributing $X$ according to $\nu$ at time $T+m$, we set $T(1) = T+m$. 
Otherwise, we continue simulating $X$ forward from time $T+m$, and continue attempting to distribute $X$ according to $\nu$ at successive visits to $A$ until we are successful, thereby defining the first regeneration time $T(1)$. 
We then successively follow the same procedure from time $T(1)$ forward to construct $T(2),T(3),\ldots$. 
The Markov chain $X$ is \emph{wide-sense regenerative} with respect to the sequence of random times $T(1),T(2),\ldots$. 
In particular, the random element $((X_{T(i)+j}, T(i+j+1)-T(i)): j\geq 0)$ is identically distributed and independent of $T(i)$ for $i\geq 1$; see \cite{Meyn2009} and \cite{Thorisson2000} for details. As noted in \cite{Asmussen2007}, one can implement the above algorithm using acceptance / rejection so that explicit generation from the conditional distribution (given $X_T$ and $X_{T+m}$) can be avoided. 

We now explain our exact estimation algorithm in the special case that $X$ is aperiodic, and we later generalize to the periodic case. 
As in Section~\ref{sec:2}, the key is to construct a coupling of $(X_{n-1}, X_n)$ that makes $\Delta_n$ small. 
Specifically, we will construct alongside $(X_n:n\geq 0)$ another sequence $(X_n': n\geq 0)$ such that $(X_n': n\geq 0)\stackrel{\mathcal D}{=} (X_n : n \geq 0)$, and then attempt to (distributionally) couple the $X_n'$'s to the $X_{n+1}$'s so that $X_{\tau} \stackrel{\mathcal D}{=}X'_{\tau-1}$.

We start by drawing $X_0$ from the distribution $\nu$ and set $X_0' = X_0$. 
We have already discussed the simulation of $X$ and the construction of the associated $T(n)$'s. 
Conditional on $X_0$, we simulate $(X_n': n \geq 1)$ independently of $(X_n: n\geq 1)$, thereby producing an associated sequence of regeneration times $0 = T'(0) < T'(1) < T'(2) < \ldots$. 
We then let the (distributional) coupling time $\tau$ be the first time at which one of the $(T(j)-1)$'s coincides with one of the $T'(i)$'s, specifically $\tau = \inf\{T(n): n\geq 1, \text{ there exists } m \geq 0 \text{ such that } T'(m) = T(n)-1\}$. 
With this definition of $\tau$, $(X_{\tau+j}: j\geq 0) \stackrel{\mathcal D}{=} (X'_{\tau+j-1} : j \geq 0)$ and $((X_{\tau+j}, X'_{\tau+j-1}): j \geq 0)$ is independent of $\tau$. 
As a consequence, if we set $\Delta_k = (f(X_k) - f(X_{k-1}'))\I(\tau > k)$ (with $f$ bounded), it follows that 
	\begin{align*} 
	\E \Delta_k
	&= \E\left(f(X_k)-f(X_{k-1}')\right)\I(\tau>k)\\
	&= \E\left(f(X_k)-f(X_{k-1}')\right)\I(\tau>k)\\
	&\quad + \sum_{j=1}^k \E \left(f(X_{\tau+k-j})-f(X_{\tau+k-j-1}')\right)\P(\tau = j)\\
	&= \E\left(f(X_k)-f(X_{k-1}')\right)\I(\tau>k)\\ 
	&\quad + \sum_{j=1}^k \E \left(f(X_{\tau+k-j})-f(X_{\tau+k-j-1}')\right)\I (\tau = j)\\
	&= \E\left(f(X_k)-f(X_{k-1}')\right) = \E\big(f(X_k)-f(X_{k-1})\big).
	\end{align*}
Furthermore, the aperiodicity of $X$ and the boundedness of $f$ imply that $\E f(X_k) \to \E f(X_\infty)$, where $X_\infty$ has the distribution of $\pi$, with $\pi$ being the unique stationary distribution of the Harris chain.

It remains to establish condition (\ref{eq:2.2}). 
Observe that the $(T(n)-1)$'s are the regeneration times for the sequence $(X_{n+1}: n \geq 0)$, in which $X$ is initialized with the distribution $\P(X_1 \in \cdot)$. 
Equivalently, the $(T(n)-1)$'s are renewal times for the delayed renewal process where the inter-renewal times share the same inter-renewal distribution as for the $T'(n)$'s, but in which the probability mass function for the initial delay is given by $(q_j: j \geq 0)$, where $q_j = \P(T'(2) - T'(1) = j+1)$. 
Thus, $\tau$ is the first time that two independent aperiodic renewal processes couple, in which one is non-delayed (corresponding to $(X_n': n \geq 0)$) and the other is delayed with initial delay $(q_j: j \geq 0)$. 
According to \cite{Lindvall2002}, p.27, $\E \tau^r<\infty$ for $r\geq 1$, provided that $\E(T(2) - T(1))^r < \infty$. 
(Note that $\sum_{j=0}^\infty j^r q_j < \infty = \E (T(2) - T(1))^r$ for the specific delay distribution that arises here.) Of course, the positive recurrence of $X$ implies that $\E (T(2)-T(1))<\infty$ (see \cite{Athreya1978}), implying that $\E \tau < \infty$. 
Hence
	\begin{equation*}
	\E \sum_{k=0}^\infty |\Delta_j| \leq 2\,\|f\|\, \E \sum_{k=0}^\infty \I(\tau>k) = 2 \|f \|\, \E \tau < \infty,
	\end{equation*}
where $\|f\| = \sup\{|f(x)|: x \in S\}<\infty$, validating (\ref{eq:2.2}). 
It follows that $Z$ is an unbiased estimator for $\E f(X_\infty)$.

For the periodic case (with period $p$), we can apply the above algorithm to $(X_{pn}: n \geq 0)$, and apply the coupling $\tau$ to coupling the $X_{pn}$'s to the $X_{p(n-1)}$'s. 
(Note that by setting $X_0 = X_0'$, we guarantee that both $(X_n: n \geq 0)$ and $(X_n': n \geq 0)$ start off in the same periodic sub-class, so that $(X_{p(n+1)}: n \geq 0)$ can successfully couple with $(X'_{pn}: n \geq 0)$.) We summarize our discussion thus far with the following result. 

\begin{thm}\label{thm:2} If $X$ is a positive recurrent Harris chain and $f$ is bounded, the estimator $Z$ described above is unbiased for $\E f(X_\infty)$.
\end{thm} 

Of course, this estimator may fail to exhibit a ``square root convergence rate,'' because $Z$ may not have finite variance and the expected computation time to generate $Z$ may be infinite. 
However, we note that because $\Delta_k = 0$ for $k > \tau$ in this setting, the number of time steps of $((X_j, X_j'): j \geq 0)$ that need to be simulated in order to compute $Z$ is bounded by $2\min(\tau, N)$. 
(The factor of 2 appears because we need to simulate both the $X_j$'s and $X_j'$'s.) 
Hence, if $\xi$ is a measure of the computational effort required to generate $Z$, $\E \xi$ is automatically finite because $\E \tau < \infty$, regardless of the distribution of $N$. 
(In fact, we may set $N=\infty$ a.s. in this setting, if we so wish.)

Turning now to the variance of $Z$, we note that if $f$ is bounded, $\E \Delta_k^2 = O(\P(\tau>k))$ as $k\to  \infty$. Furthermore, if $\E (T(2)-T(1))^r < \infty$ for $r>1$, then $\E\tau^r < \infty$, so that $\E \Delta_k^2 = O(k^{-r})$ by virtue of Markov inequality. Also, for $k<j$, $\Delta_k \Delta_j = 0$ unless $\tau> j$, so $\E \Delta_k \Delta_j = O(j^{-r})$ as $j \to \infty$, uniformly in $k$. Thus
	\begin{equation}
	\sum_{k=j+1}^\infty \E \Delta_k \Delta_j = O(k^{1-r})
	\end{equation}
as $k\to\infty$. In order that there exist a probability distribution $N$ so that $\E Z^2 < \infty$, (\ref{eq:*}) implies that it is therefore sufficient that $\alpha > 2$ (in which case we can, for example, choose $N$ so that $\P(N\geq k)$ is of order $k^{1-\alpha/2}$ for $k$ large).

We have therefore proved the following theorem, establishing a ``square root convergence'' (in the computational effort $c$) for our estimator.

\begin{thm}\label{thm:3.2}
If $X$ is Harris chain with $\E(T(2)-T(1))^r < \infty$ for $r>2$ and $f$ is bounded, then
\begin{equation*}
c^{1/2}(\bar Z_{\Gamma(c)} - \E f(X(\infty))) \Rightarrow \sqrt{\E \xi\cdot \var Z} \ N(0,1)
\end{equation*}
as $c\to \infty$.
\end{thm}

An improvement to the above coupling can be easily implemented. 
In the above algorithm, $\tau$ occurs whenever $X$ and $X'$ $m$ time steps earlier were in $A$, and both $X$ and $X'$ independently chose at that time to distribute themselves according to $\nu$ $m$ time units later. 
But an alternative coupling is to generate $(X_{T+m}, X'_{T+m-1})$ as follows, whenever $(X_T, X'_{T-1})\in A\times A$. 
As in the previous algorithm, we distribute $X_{T+m}$ according to $\nu$ with probability $\lambda$, and according to $Q(X_T, \cdot)$ with probability $1-\lambda$. 
However, we now modify the dynamics for $X'$. 
Whenever $X_{T+m}$ is distributed according to $\nu$, set $X'_{T+m-1} = X_{T+m}$. 
On the other hand, whenever $X_{T+m}$ is distributed according to $Q(X_T, \cdot)$, independently generate $X'_{T+m-1}$ according to $Q(X'_{T-1}, \cdot)$. 
This coupling preserves the marginal distribution of $X$ and $X'$, but the time $\tau'$ at which $X$ and $X'$ couple (so that $X_{\tau'}=X'_{\tau'-1}$) is a.s.\ smaller than under the previous ``independent coupling.'' Consequently, $\P(\tau' \geq k) \leq \P(\tau\geq k)$ for $k\geq 0$, so $\E (\tau')^r \leq \E \tau^r$ for $r>0$, thereby establishing that this coupling can be used in place of $\tau$ in proving Theorem~\ref{thm:3.2}. Given that $\tau' \leq \tau$, this coupling is computationally preferable to $\tau$.

%%%%%%%%%%%%%%%%%%%%%%%%%%%%%%%%%%%%%%%%%%%%%%%%%%%%%%%%%%%%%%%%%%%%%%%%%%%%
%
%
% Section 4. A Glivenko-Cantelli Result
%
%
%%%%%%%%%%%%%%%%%%%%%%%%%%%%%%%%%%%%%%%%%%%%%%%%%%%%%%%%%%%%%%%%%%%%%%%%%%%%
\section{A Glivenko-Cantelli Result}\label{sec:4}
In some settings, one may be interested in computing the equilibrium distribution of some real-valued functional $f$ of the Markov chain, rather then merely its expected value $\E f(X_\infty)$. In this section, we study the behavior of our unbiased estimator for the equilibrium probability $\P(f(X_\infty) \leq x)$ ($= \E \I(f(X_\infty)\leq x)$) as  a function of $x$. Because the mapping $\I(f(\cdot)\leq x)$ is not Lipschitz, Section~\ref{sec:2}'s theory does not apply. As a consequence, we focus here on the case where $X$ is a positive recurrent Harris chain.

Set $Y_k = f(X_k)$ and $Y_k' = f(X_k')$. Let $((Y_{k,j}, Y_{k,j}'): 0 \leq k \leq \min(\tau_j, N_j)): j \geq 1)$ be a sequence of iid copies of $((Y_k, Y_k'): 0 \leq k \leq \min(\tau, N))$, where the $Y_k$'s and $Y_k'$'s are constructed as in Section~\ref{sec:3}. The empirical measure (intended to estimate $F_\infty(\cdot)\triangleq \P(f(X_\infty)\leq \cdot)$) associated with sample size $n$ is then given by the random signed measure
	\begin{equation*}
	\pi_n(\cdot) = \frac{1}{n}\sum_{j=1}^n \sum_{k=0}^{\tau_j \wedge N_j} \frac{\delta_{Y_{k,j}}(\cdot) - \delta_{Y_{k-1,j}'}(\cdot)}{\P(N\geq k)},
	\end{equation*}
where $a\wedge b \triangleq \min(a,b)$, and $\delta_y(\cdot)$ is a unit point mass measure at $y$. Observe that 
	\begin{equation*}
	\int_S y \pi_n(dy)
	\end{equation*}
is Section~\ref{sec:3}'s unbiased estimator for $\E f(X_\infty))$, and 
	\begin{equation*}
	F_n(x) \triangleq \int_S \I(y \leq x) \pi_n(dy)
	\end{equation*}
is Section~\ref{sec:3}'s unbiased estimator for $F_\infty(x)$. We may re-write $F_n(\cdot)$ as 
	\begin{equation*}
	F_n(x) = \sum_{k=0}^\infty \frac{1}{n} \sum_{j=1}^n (\I(Y_{k,j} \leq x) - \I(Y'_{k-1,j} \leq x)) \frac{\I(\tau_j \wedge N_j \geq k)}{\P(N\geq k)}.
	\end{equation*}
Because the sample functions
	\begin{equation*}
	\frac{1}{n} \sum_{j=1}^n \I(Y_{k,j} \leq x) \I(\tau_j \wedge N_j \geq k)
	\end{equation*}
and
	\begin{equation*}
	\frac{1}{n} \sum_{j=1}^n \I(Y_{k-1,j}' \leq x)\I(\tau_j \wedge N_j \geq k)
	\end{equation*}
are monotone in $x$, a proof identical to that of the standard Glivenko-Cantelli theorem (see, for example, \cite{Chung2001}) establishes that 
	\begin{equation*}
	\sup_x \left| \sum_{j=1}^n\I(Y_{kj}\leq x) \I(\tau_j\wedge N_j \geq k) - \E \I(Y_k \leq x) \I(\tau\geq k) \P(N_j \geq k) \right| \to 0 \quad a.s.
	\end{equation*}
and 
	\begin{equation*}
	\sup_x \left| \sum_{j=1}^n\I(Y_{k-1,j}'\leq x) \I(\tau_j\wedge N_j \geq k) - \E \I(Y'_{k-1} \leq x) \I(\tau\geq k) \P(N_j \geq k) \right| \to 0 \quad a.s.
	\end{equation*}
as $n \to \infty$, for each fixed $k\geq 0$. Since we proved in Section~\ref{sec:3} that 
	\begin{equation*}
	\E (\I(Y_k \leq x) - \I(Y_{k-1}' \leq x)) \I(\tau\geq k) = \P(Y_k \leq x) - \P(Y_{k-1} \leq x),
	\end{equation*}
it follows that for any $m\geq 1$, 
	\begin{equation}\label{eq:4.1}
	\sup_x \left| \sum_{k=0}^m \frac{1}{n} \sum_{j=1}^n (\I(Y_{kj}\leq x) - \I(Y_{k-1,j}'\leq x))\frac{\I(\tau_j \wedge N_j \geq k)}{\P(N\geq k)} - \P(Y_m \leq x) \right| \to 0 \quad a.s.
	\end{equation}
as $n\to \infty$. If $X$ is any aperiodic positive recurrent Harris chain, $Y_m$ converges to $Y_\infty$ in total variation, and hence
	\begin{equation}\label{eq:4.2}
	\sup_x |\P(Y_m \leq x) - F_\infty(x) | \to 0
	\end{equation}
as $m \to \infty$. (If $X$ is periodic, we adapt (\ref{eq:4.2}) by restricting $m$ to multiples of $p$, and (\ref{eq:4.2}) is replaced by 
	\begin{equation*}
	\sup_x \left|\frac{1}{p}\sum_{i=0}^{p-1}\P(Y_{m+i} \leq x) - F_\infty(x) \right| \to 0
	\end{equation*}
as $m \to \infty$.)

Suppose now that $\P(N\geq k ) \sim ck^{-\alpha}$ as $k\to \infty$ (where $a_k \sim b_k$ means that $a_k / b_k \to 1$ as $k \to \infty$). Note that for $m$ sufficiently large, 
	\begin{align}\label{eq:4.3}
	&\sum_{k > m} \left| \frac{1}{n} \sum_{j=1}^n (\I(Y_{kj} \leq x) - \I(Y_{k-1,j}' \leq x) ) \frac{\I(\tau_j \wedge N_j \geq k)}{\P(N\geq k)}\right| \nonumber\\
	&\leq \sum_{k > m} \frac{1}{n} \sum_{j=1}^n \frac{\I(\tau_j \wedge N_j \geq k)}{\P(N\geq k)} \nonumber\\
	&\leq \frac{2}{c} \sum_{k > m} k^\alpha \frac{1}{n} \sum_{j=1}^n \I(\tau_j \wedge N_j \geq k) \nonumber\\
	&\leq \frac{3}{c(\alpha+1)} \cdot \frac{1}{n} \sum_{j=1}^n (\tau_j \wedge N_j )^{\alpha+1} \I(\tau_j \wedge N_j \geq m) \nonumber\\
	&\to \frac{3}{c(\alpha+1)} \E (\tau \wedge N)^{\alpha+1} \I(\tau \wedge N \geq m) \qquad a.s.
	\end{align}
as $n \to \infty$. If $\E (\tau\wedge N)^{\alpha + 1} < \infty$, it follows that 
	\begin{equation}\label{eq:4.4}
	\E (\tau\wedge N)^{\alpha+1} \I(\tau\wedge N \geq m) \to 0
	\end{equation}
as $m \to \infty$. 

By first fixing $m$, then letting $n \to \infty$, and finally sending $m \to \infty$, (\ref{eq:4.1})-(\ref{eq:4.4}) therefore prove that 
	\begin{equation*}
	\sup_x |F_n (x) - F_\infty(x) | \to 0
	\end{equation*}
a.s. as $n\to \infty$. It remains to consider the finiteness of $\E (\tau\wedge N)^{\alpha+1}$. Observe that 
	\begin{equation*}
	\E (\tau\wedge N)^{\alpha+1} \leq (\alpha+1)\sum_{k=0}^\infty k^\alpha \P(\tau\geq k) \P(N\geq k). 
	\end{equation*}
But $\sum_{k=0}^\infty \P(\tau\geq k) = \E (\tau + 1) < \infty $ and $k^{\alpha} \P(N\geq k) \to c$ as $k \to \infty$, thereby proving that $\E(\tau\wedge N)^{\alpha+1}$ is automatically finite. 

This proves the following Glivenko-Cantelli type theorem for the estimator of Section~\ref{sec:3}. 

	\begin{thm}\label{thm:4.1}
	Suppose $\P(N\geq k)\sim ck^{-\alpha}$ as $k \to \infty$ for $\alpha > 0$. If $X$ is a positive recurrent Harris chain, then $F_{n}(x)$ is an unbiased estimator for $F_\infty(x)$ for each $x\in \R$, and 
		\begin{equation*}
		\sup_x|F_n(x)  - F_\infty(x)| \to 0\qquad a.s.
		\end{equation*}
	as $n \to \infty$. 
	\end{thm}

%%%%%%%%%%%%%%%%%%%%%%%%%%%%%%%%%%%%%%%%%%%%%%%%%%%%%%%%%%%%%%%%%%%%%%%%%%%%
%
%
% Section 5. Numerical Results
%
%
%%%%%%%%%%%%%%%%%%%%%%%%%%%%%%%%%%%%%%%%%%%%%%%%%%%%%%%%%%%%%%%%%%%%%%%%%%%%
\section{Numerical Results}\label{sec:5}
We present here a brief account of the numerical performance of our exact estimation algorithms. For our contracting chain example, we consider the non $\phi$-irreducible Markov chain $X=(X_n: n\geq 0)$ given by
$$ X_{n+1} = \frac{1}{2} X_n + V_{n+1},$$
where the $V_i$'s are iid with $\P(V_n = 0) = 1/2 = \P(V_n = 1)$, with corresponding Lipschitz functions $f_1(x) = x$, $f_2(x) = \min(1, x)$, and $f_3 (x) = x^2$. For this example, $\pi$ is uniform on $[0,2]$, $\E f_1(X_\infty) =1$, $\E f_2(X_\infty) = 3/4$, and $\E f_3(X_\infty) = 4/3$. Tables~\ref{table:contracting} and \ref{table:contracting_large_mean} report results for two different distributions for $N$. As expected, the algorithm based on $Z$ becomes more attractive when $N$ has a heavier tail, because the computational effort for $Z^*$ increases quadratically in $N$ (because of the non-recursive computation of the $\Delta_i$'s), whereas the effort for $Z$ increases linearly in $N$.

\begin{table}
\center

\caption{ Contracting chain: with $10^6$ time steps, $X_0 = 1$, $\P(N\geq n) = 2^{1-n}$ \label{table:contracting}}
\begin{tabular}{c|c|c|c}
\hline

$f(x)$ & Estimator & 90\% Confidence Interval& \#Samples 
\\\hline

$x$
& $Z$\hspace{3pt} & 1.013\hspace{1\zero} $\pm$ 1.1$\times 10^{-2}$ & 6.7$\times 10^{4}$\\
& $Z^*$           & 0.9974\hspace{0\zero} $\pm$ 7.3$\times 10^{-3}$ & 5.0$\times 10^{4}$\\\hline

$\min(1,x)$
& $Z$\hspace{3pt} & 0.7531\hspace{0\zero} $\pm$ 6.2$\times 10^{-3}$ & 6.7$\times 10^{4}$\\
& $Z^*$           & 0.7552\hspace{0\zero} $\pm$ 4.7$\times 10^{-3}$ & 5.1$\times 10^{4}$\\\hline

$x^2$
& $Z$\hspace{3pt} & 1.344\hspace{1\zero} $\pm$ 2.3$\times 10^{-2}$ & 6.7$\times 10^{4}$\\
& $Z^*$           & 1.334\hspace{1\zero} $\pm$ 1.6$\times 10^{-2}$ & 5.0$\times 10^{4}$\\\hline

\end{tabular}

\vspace{10pt}

\caption{ Contracting chain: with $10^6$ time steps, $X_0 = 1$, $\P(N\geq n) = 0.95^{n-1}$ \label{table:contracting_large_mean}}
\begin{tabular}{c|c|c|c}
\hline

$f(x)$ & Estimator & 90\% Confidence Interval & \#Samples
\\\hline

$x$
& $Z$\hspace{3pt} & 1.009\hspace{1\zero} $\pm$ 3.3$\times 10^{-2}$ & 2.7$\times 10^{3}$\\
& $Z^*$           & 1.006\hspace{1\zero} $\pm$ 6.1$\times 10^{-2}$ & 2.4$\times 10^{2}$\\\hline

$\min(1,x)$
& $Z$\hspace{3pt} & 0.743\hspace{1\zero} $\pm$ 1.7$\times 10^{-2}$ & 2.7$\times 10^{3}$\\
& $Z^*$           & 0.764\hspace{1\zero} $\pm$ 3.6$\times 10^{-2}$ & 2.5$\times 10^{2}$\\\hline

$x^2$
& $Z$\hspace{3pt} & 1.356\hspace{1\zero} $\pm$ 6.7$\times 10^{-2}$ & 2.7$\times 10^{3}$\\
& $Z^*$           & 1.39\hspace{2\zero} $\pm$ 1.3$\times 10^{-1}$ & 2.5$\times 10^{2}$\\\hline

\end{tabular}

\end{table}

We turn next to the Harris chain algorithm, implemented with the coupling $\tau'$ of Section~\ref{sec:3}. We consider the Markov chain $W=(W_n: n\geq 0)$ on $\R_+$ corresponding to the waiting time sequence for the $M/M/1$ queue, with arrival rate 1/2 and unit service rate. The equilibrium distribution $\pi$ here is a mixture of a unit point mass on 0 and an exponential distribution with rate parameter $1/2$, with equal probability 1/2 on each mixture component. We let the function $f$ be given by $f(x) = \I(x > 1)$, so that $\P( W_\infty > 1) = \frac{1}{2} e^{-1/2} \approx 0.303.$

As for $N$, we note that the same proof technique as for Proposition 1 of \cite{Rhee2013} establishes that the optimal choice for the distribution of $N$ is to choose $\P(N\geq k)$ proportional to 
	\begin{equation*}
	\sqrt{\frac{\E \Delta_k^2 + 2\sum_{j = k+1}^\infty \E \Delta_k \Delta_j}{\P(\tau \geq k)}},
	\end{equation*}
provided that this sequence is non-\revision{increasing}. Since it seems likely that $\E \Delta_k^2$ will frequently be of roughly the same order as $\P(\tau \geq k)$ for large $k$, this suggests that the optimal distribution will often have positive mass at infinity. In view of this observation, we have chosen to use a very heavy-tailed specification for $N$, namely $\P(N\geq k)= 1/k$ for $k\geq 0$. Table~\ref{table:HarrisChain} reports the results of our computations with $\lambda=1$ and with small set $A=\{0\}$; the results show the ``square root'' decrease in the width of the confidence interval that is to be expected.

\begin{table}\center
\caption{ Harris chain  \label{table:HarrisChain}}
\begin{tabular}{c|c|c}\hline

\#Steps Simulated  & 90\% Confidence Interval & \#Samples \\\hline

1.0$\times 10^{5}$ & 0.283\hspace{1\zero} $\pm$ 8.3$\times 10^{-2}$ & 3.2$\times 10^{4}$\\

2.0$\times 10^{5}$ & 0.279\hspace{1\zero} $\pm$ 5.8$\times 10^{-2}$ & 6.5$\times 10^{4}$\\

5.0$\times 10^{5}$ & 0.296\hspace{1\zero} $\pm$ 3.4$\times 10^{-2}$ & 1.6$\times 10^{5}$\\

1.0$\times 10^{6}$ & 0.329\hspace{1\zero} $\pm$ 2.7$\times 10^{-2}$ & 3.2$\times 10^{5}$\\

2.0$\times 10^{6}$ & 0.294\hspace{1\zero} $\pm$ 1.8$\times 10^{-2}$ & 6.5$\times 10^{5}$\\

5.0$\times 10^{6}$ & 0.308\hspace{1\zero} $\pm$ 1.2$\times 10^{-2}$ & 1.6$\times 10^{6}$\\

1.0$\times 10^{7}$ & 0.2992\hspace{0\zero} $\pm$ 8.4$\times 10^{-3}$ & 3.2$\times 10^{6}$\\

2.0$\times 10^{7}$ & 0.3089\hspace{0\zero} $\pm$ 5.8$\times 10^{-3}$ & 6.5$\times 10^{6}$\\

5.0$\times 10^{7}$ & 0.2995\hspace{0\zero} $\pm$ 3.7$\times 10^{-3}$ & 1.6$\times 10^{7}$\\

1.0$\times 10^{8}$ & 0.3041\hspace{0\zero} $\pm$ 2.6$\times 10^{-3}$ & 3.3$\times 10^{7}$\\

2.0$\times 10^{8}$ & 0.3024\hspace{0\zero} $\pm$ 1.9$\times 10^{-3}$ & 6.5$\times 10^{7}$\\

5.0$\times 10^{8}$ & 0.3036\hspace{0\zero} $\pm$ 1.2$\times 10^{-3}$ & 1.6$\times 10^{8}$\\

\hline
\end{tabular}
\end{table}

\section*{Acknowledgements}
The authors gratefully acknowledge the support of National Science Foundation Grant DMS-1320158 and a Samsung Scholarship.
% Place the text of your acknowledgements after the \acks command.
% \acks generates the heading "Acknowledgements".
% If you wish to make only one acknowledgement, use \ack.
% \ack generates the heading "Acknowledgement".

% Reference list
%
% References should be in the following form (or the BibTeX file
% apt.bst should be used):
%
% For a journal:
% Surname, Initial (year). Title of paper. {\em Journal title}
% {\bf Vol,} page--range.
%
% For a book:
% Surname, Initial (year). {\em Book title}. Publisher, Address.
%
% Note the following example of a reference list.

% \begin{thebibliography}{99}
% \footnotesize

% \bibitem{ref1}
% {\sc Ball, K. and Chain, H.} (1988). {\em Kurtosis: A Critical
% Review}, 2nd~edn. John Wiley, New York.

% \bibitem{ref2}
% {\sc Boyd, W.} (1978). Hyperbolic distributions. Doctoral Thesis,
% University of Boston School of Mathematics.

% \bibitem{ref3}
% {\sc Sichel, H.~S., Kleingeld, W.~J. and Assibey-Bonsu, W.}
% (1992).  A comparative study of three frequency-distribution
% models for use in ore valuation. {\em J. S. Afr. Inst. Min. Met.}
% {\bf 92,} 91--99.

% \end{thebibliography}
% \bibliographystyle{apt}
\bibliographystyle{apalike}
\bibliography{References-ExactEstimation}

\end{document}